Protein Folding and Machine Learning: Fundamentals

Walter A. Simmons

October 2018

Department of Physics and Astronomy

University of Hawaii at Manoa

## Abstract

In spite of decades of research, much remains to be discovered about folding: the detailed structure of the initial (unfolded) state, vestigial folding instructions remaining only in the unfolded state, the interaction of the molecule with the solvent, instantaneous power at each point within the molecule during folding, the fact that the process is stable in spite of myriad possible disturbances, potential stabilization of trajectory by chaos, and, of course, the exact physical mechanism (code or instructions) by which the folding process is specified in the amino acid sequence.

Simulations based upon microscopic physics have had some spectacular successes and continue to improve, particularly as super-computer capabilities increase. The simulations, exciting as they are, are still too slow and expensive to deal with the enormous number of molecules of interest.

In this paper, we introduce an approximate model based upon physics, empirics, and information science which is proposed for use in machine learning applications in which very large numbers of sub-simulations must be made. In particular, we focus upon machine learning applications in the learning phase and argue that our model is sufficiently close to the physics that, in spite of its approximate nature, can facilitate stepping through machine learning solutions to explore the mechanics of folding mentioned above.

We particularly emphasize the exploration of energy flow (power) within the molecule during folding, the possibility of energy scale invariance (above a threshold), vestigial information in the unfolded state as attractive targets for such machine language analysis, and statistical analysis of an ensemble of folding micro-steps.

# INTRODUCTION

In spite of more than half a century of intensive research, the fundamental mechanisms of protein folding are still uncertain; the 'code' that carries the folding instructions in the amino acid sequence is still not known; in fact, even the most elementary physical structures that carry the instructions are not entirely agreed upon. Since protein folding is a self-organizing process, the common conceptualization of a control function and a state space is complicated. Furthermore, with ten degrees of freedom per residue, identification of a simple subset, such as dihedral angles only, is questionable. Finally, many folding processes occur near the limit of thermodynamic instability.

In short, the simulation of many proteins has to manage against the accumulation of errors while maneuvering a low energy signal, over a long time frame, through a noisy background based upon uncertain parameters (1), (2).

However, proteins are not disrupted by small changes in initial conformation, in temperature, or in chemical potential with the solution, and fold quickly to unique final states. Evidently, natural molecules have some aspects that make small changes in ambient conditions unimportant. (3), (4), (5), (6).

Simulation is the most common approach to determining the ground state structure from the sequence. In this paper, we discuss the foundations of an alternate approach using machine learning (ML).

There have been many successes with simulations, but a major breakthrough will be needed to increase the speed and lower the cost, so that the structure of the tens of thousands (or more) medically important molecules can be obtained by computation alone.  One major limitation of simulation, by which we mean computer simulations built upon physics and chemistry based models of the molecules, is the fact that simulations in femtosecond time intervals might require $10^{10}$ or more time intervals.  The models use changing physical parameters obtained from models or from chemistry experiments with finite precision.  Also, the natural instruction set has some (unknown) finite precision.  The undefined parts of the instructions, whatever that might mean, enter in a non-linear way. Therefore, these limitations taken together or separately, imply that accumulative errors may develop and invalidate the simulation.  This is typically treated by intensive computation, (such as minimizing the length of the time step), which is a substantial time and cost burden.

The acid test of a simulation is the final state calculation; however, we propose that much can be learned using ML even if the detailed ground state cannot be calculated precisely.

A final remark on simulations.  The unfolded, or initial state, is only partially known.  It may be the instructions essential to folding are partially encoded in conformations of this state, over and beyond the sequence (7).  If so, a model that is tested against randomly constructed conformations may fail in some cases even if the instructions inherent in the sequence are properly accounted for.

The dramatic success of AlphaGo-Zero in solving Go (8) has undoubtedly set the imaginations of protein scientists aflame.  ML has been applied in many ways; some interesting applications: (9), (10), (11), (12), (13) {High energy physicists have also found

Artificial Intelligence (AI) to be a set of powerful tools (14), (15), (16), (17), (18).} This paper explores the folding fundamentals upon which machine-learning models might be built. An obvious advantage of ML over simulation is that a constant looping on a known structure (in the machine learning phase) can, in principle, reduce or eliminate accumulative calculational errors.

In contemplating the application of machine learning, a statistical matrix is often adjusted in a loop to achieve pattern recognition. The fact that many valid pathways may pass near thermodynamic instabilities means that many pathways derived in the machine learning would not be valid in nature.

Therefore, eliminating accumulated errors inherent in simulations, eliminating unphysical trajectories, and accommodating regions of energy scale invariance are desirable features of computer based folding.

From information theory, (19), (20), (21), we note that if the symbols used in communication are energy-symmetric (same energy to make and use a zero as a one), then the message need not be modulated in energy. We discuss, below, how that might appear in folding as, 'thermodynamic perfection'.

Also from information theory, we have the concept that the distribution function of the set of all possible messages with a given format and a common reservoir of symbols with fixed probabilities, is a Gaussian. Furthermore, the number of messages with the highest probabilities are those beneath the top of the Gaussian (i.e. within the variance). This useful approach has been applied to molecular biology and we discuss how this might be useful in folding research in particular.

We summarize our objectives in this paper as follows.

Develop a model of folding that is simple but which respects the physics and especially the geometry of the molecules. Base the model upon a probabilistic process (e.g. directed Brownian motion); the process will define an ensemble of microscopic folding steps based upon informational 'words'. Propose ML in the learning phase to eliminate accumulation errors and to accommodate the weak ambient energy dependence of the final state structure. Consider an ensemble of molecules and show how the 'words' close to the center of the Gaussian can be expected to dominate; this enables immediate application of the maximum entropy method of pattern resolution. We shall not enter into the specifics of algorithms (22), which can be written in many ways. We are attempting to lay a groundwork based on empirical knowledge and physics, which can be used in variously in algorithms.

Parameterization

As mentioned, there is an open question of which physical proprieties of the sequence carry the instructions and exactly how that works.

It is useful to consider the following generic description. The folding consists of two major processes: a random Brownian component and a directed part driven by the (limited) energy initially resident in the unfolded molecule (23). The directed force takes the specific form of driven torsion waves. The wave motion is enormously complicated since the boundary conditions change continuously during folding due to flexing of the molecule and change discontinuously due to contact formation.

Rotations in general do not commute and the fact the folding goes to a unique native structure, strongly suggests that there are few alternative directed paths.

It is worth remarking that it is well-known in physics (23), (24), (25), that stochastic forces can smooth out perturbations in addition to driving a system down-hill. Whether that occurs in folding is not known.

The use of a standard parameterization used in simulations is not particularly attractive for the ML applications envisioned here. Instead, we propose an informational parameterization based upon microscopic changes in angles consistent with molecular geometry. Detailed models can be exploited later.

In this paper, we attack the first step in ML learning about folding. We parameterize fundamental units of instructions that can change on the computer without taking the model into unphysical territories.

We begin by recognizing that changes in shape across some range of residues is fundamental; moreover, rotations are the most important changes, so the chain must encode directions as well as magnitudes. We collect together pieces of various models, all well known.

In the earliest days of folding simulations, attention turned to triplets of residues; that is, one residue and two nearest neighbors. It is now understood that coding instructions apply over domains, if not over larger structures, and not just locally around such a triplet. In spite of these limitations, we suggest the triplet, a change in which we shall call a 'word' of instruction, be used in ML designs.

The major reasons for this choice are as follows.

a.) With three dihedral angles per residue, a triplet of adjacent residues has six degrees of freedom. That is just as required to describe the relative positions (three) and orientations (three) of the ends of the triplet.

b.) The instructions in the sequence change during folding. The angular change and energy change can be accommodated in a triplet.

c.) The changes in the triplet are, in part, associated with the deterministic gross motion and, in part, by random Brownian effects. The random part is geometrical and probabilistic and can be treated without reference to energy; i.e. non energy directed motion must be included in the computer calculation.

d.) Except for steric hindrance, the triplet can take the shape of any structural element. It is also flexible enough to traverse the passage through the exit of the ribosome.

e.) Because of the short length, inevitable computation errors can be controlled over short distances.

f.) The short word length is convenient for analysis of wave motion and is especially convenient for studying the change in wave structure resulting from contact formation.

g.) Power: success in seeing folding take place step by step would create opportunities to study instantaneous power during folding. Note that some misfolds, or alternate viable conformations, are blocked by local power limitations even while the average power may be sufficient to overcome

> some local thresholds. The importance of such detailed understanding was emphasized by Dobson in 2003 (7). See (26) for recent and detailed discussion.

The unfolded state must also receive some attention. As mentioned, some unrecognized aspects of this collection of states may define part of the folding process. To carry that one step further, we suggest that the various microscopic features follow probability distribution functions. In that case, as we discussed in another context, the Central Limit Theorem suggests that these instructional features follow a Gaussian and only a microscopic number of possible such features are actually in use. In other words, the instructions in the amino acid sequence also specify the unfolded state and a Gaussian form suggests that the number of unfolded states is likely very small.

This can be conceptualized by visualizing energy flowing through a limited number of Levinthal pathways; the power may originate in a highly non-uniform distribution of initial energy in the unfolded sates.

In initial development the energy dependence of the word can be simplified because the word is essential geometric and intended to have sufficient flexibility to fit nearly any chain shape. When the machine learns enough to require an energy model that can be inserted at that phase.

## Model Construction and Goals

As mentioned, we follow information theory to develop an analysis that is broader than statistical mechanics in that changes the do not depend upon the energy stored in the initial state are included.

The basic 'word' of code instruction embodied in the sequence is, according to the ideas just introduced, is a six-dimensional differential, (or difference), change in the six dihedral angles of the triplet.

Contact formation must be modeled by computing the position of the chain, detecting contact, and making an evaluation of binding probability.

Connecting the words of the sequence introduces a problem common to all computer simulations: the steps are carried out individually and have to be smoothed out.  The ML approach is much less sensitive to this problem than strictly model based simulations.

There are various statistical methods that follow along the lines just described.  A newly created one can be found at (14), which is a physics application.

Since we have referred to the common approach of treating folding as a deterministic process driven down slope by random impulse, we now analyze the proposed model with that in mind.

To begin, we note that if the machine learns to fold specific structures, then one can test the effect of a change in temperature VS a change in chemical potential.  That is, the relative energy scale invariance of Brownian VS driven forces.

We consider an ensemble of ML calculations based on a single set of known structures and a wide range of computer generated initial states.  We propose that the number of occurrences of any given six-dimensional word VS the specific words.  The result will consist of two parts.  The random Brownian motion will yield a Gaussian. The structure arising from the directed part will be different and we can only speculate; if it is also probabilistic (over an ensemble of folds with varying initial conditions), then it too will be Gaussian with the variance representing the most common words, as in information theory.

A related analysis is to vary the energy of the initial state. If the information theoretic model described here works, then the final structure will be independent of small changes in the initial energy (i.e. energy scale invariance). The distribution of words involved may be similar if there are a few fundamental modes of folding, or different if there are many folding pathways. If the conjecture about a non-uniform distribution of energy in the unfolded state, (mentioned above), is correct, then random initial configurations will not work.

In detail, the calculation process has a Gaussian distribution about the directed folding path. If this compares well with experiment, then it defines the folding path rather tightly.

A major point here is that the number of words appearing in a large number of folds is likely to be miniscule compared to the possible words. Said differently, certain sets of angular changes in the triplet of residues dominate over all possible sets of angles. The words that occur most frequently can be different from the words with the largest instantaneous change in energy (steepest slope).

Thus, this model, in spite of being obviously simplified, accommodates energy-directed and probabilistic changes. The model can statistically reveal the relative roles of directed and random motions as well as the presence of a few, (or of many), folding pathways.

Conclusions:

We emphasized the advantages of ML over simulations. The two most important advantages are: (i) that uncertainties in parameters and in calculations can be better suppressed in ML than in simulations and, (ii) (if only the learning phase is used, and if a sufficiently accurate model is used), then stepping through the calculation may very efficiently reveal new phenomena in folding.

We argued that correct predication of the ground state, which is the acid test of simulations, is not necessary to learn about how proteins fold. A particularly interesting short term goal is understanding instantaneous power at each point in the molecule during folding (power landscape). Some folding pathways may be allowed by available energy but forbidden by the instantaneous power requirement.

More generally, researchers frequently encounter situations in which general principles of folding might be very useful. Examples would be instantaneous power distribution, changes in the power landscape with contact formation, (proposed) energy scale invariance, underlying principles of process stability, other informational symmetry features, (e.g. such as rules for misfolding).

Dynamically, the model suggested is based closely upon the directed Brownian model and upon the idea that torsional wave motion predominates folding.

Following information theory, we introduce a 'word' describing a small change in a three-residue sub segment of the chain. We suggest that ensembles of ML tests be performed; each test has a unique initial state but all tests use the same final structure. The statistics of any probabilistic processes yield Gaussians in word frequency. The number of words found for any given molecule is expected to be minuscule compared to the number possible. We also propose searching the data for energy scale invariance.

The specific 'word' proposed is an approximation but should be very useful because of its properties listed earlier.

Finally, we remark on the extension of the application of information science to the motions involved in folding (27). If all probabilities were the same (uniform probability case) then the number of 'possible' protein sequences for a 100 residue molecule would be, as is often stated, $(20)^{100}$. However, evolution is to be viewed from here as probabilistic and the number of protein sequences that will actually occur is a microscopic fraction of the above number (20), (27). We have proposed that the same logic be applied to some defined individual angular changes during folding; the number of folding pathways is microscopic compared to the number of 'possible' pathways in the uniform probability case. [We remark that if cooperativity can be similarly limited then the number of possible folds is reduced by another exponential factor (Gaussian).] The often quoted timescale for uniform probability folding, which is $T \geq 10^{10}$ years, is also reduced exponentially by a cooperativity Gaussian and possibly another Gaussian factor due to the distribution function of energy in the unfolded state.

We believe that these calculations can be carried out using ML as described in this paper.

References:

[1] Bottaro, S.2018, & K., Lindorff-Larsen, "Biophysical experiments and biomolecular simulations: A perfect match?", Science 361, 355.

[2] Bereau, T., 2018, J.F. Rudzinski, "Circumventing additivity in molecular mechanics with conformationally-dependent surface hopping",arxiv.org/abs/1808.05644v1

[3] Levinthal, C., 1968, "Are there pathways in protein folding?" Journal de chimie physique.

[4] Dill, K.A., 1990, "Dominant Forces in Protein Folding", Biochemistry, 29(31), 7133-7155.

[5] Simmons, W. & Weiner, J.L., 2015, "Topology, Geometry, and Stability: Protein Folding and Evolution", arxiv 1505.07153.

[6] Simmons, W., 2017, "Protein Folding Problem: Scientific Basics]. Arxiv 1709.07953.

[7] Dobson, C.M., 2003, "Protein folding and misfolding", Nature 426, 884.

[8] Silver, D., 2017, J. Schrittwieser, K. Simonyan, I., Antonoglou, A.,Huang, A., Guez, , T., Hubert, L., Baker, M., Lai, A., Bolton, Y., Chen, T., Lillicrap, F., Hui, L., Sifre, G., van den Driessche, T., Graepel, D., Hassabis, "Mastering the game of Go without human knowledge", Nature 550, 354-359.

[9] Bengo, Y., 2016. Goodfellow, I.J., & Courville, A., "Deep Learning", MIT.

[10] Jurtz, V. I., 2017, A.R., Johansen, M., Nielsen, J.J.A. , Armenteros, H., Nielsen, C.K. ,Sonderby, O., Winther,  S.K., Sonderby,  "An introduction to deep learning on biological

sequence data: examples and solutions" Bioinformatics 33, 3685-3690.

[11] de Oliveira, L., Michaela Paganini, & Benjamine Nachman, "Learning Particle Physics by Example: Location-Aware Generative Adversarial Networks for Physics Synthesis"

[12] Stahl, K., Schneider, M., Brock, O., 2017,, "EPSLON-CP: using deep learning to combine information from multiple sources for protein contact prediction." BMC Bioinformatics 18, 303.

[13] Zhao, Z.Q., Zu-Gou,Y., Vo, A., Jing-Yang, W., Guo-Sheng, H. 2016, "Protein folding kinetic order prediction from amino acid sequence based on horizontal visibility network." Bioinformatics 11, 173-185.

[14] Aartsen, M.G., Ice-Cube-Gen2 Collaboration, 2018, "Computational Techniques for the Analysis of Small Signals in High-Statistics Neutrino Oscillation Experiments", arXiv: 1803.05390; pages 1-32.

[15] Baldi, P., 2014, Sadowski, P., & D. Whiteson, "Searching for exotic particles in high-energy physics with deep learning." Nature Communications volume 5, Article number: 4308.

[16] Lloyd, S., 2018 & C. Weedbrook, "Quantum Generative Adversarial Learning" PHYSICAL REVIEW LETTERS 121, 040502-1, 040502-5.

[17] Dunjko, V., 2016, J.M., Taylor, and H.J. Briegel,"Quantum Enhanced Machine Learning" PHYSICAL REVIEW LETTERS 117, 130501-1. 13050-6.

[18] Larrañaga, P., 2006, B. Calvo, R. Santana, C. Bielza, J. Galdiano, I. Inza, J.A. Lozano, R. Armananzas, G. Santafe, A. Perez, and V. Robles, "Machine learning in bioinformatics",

Briefings in Bioinformatics, Volume 7, Issue 1, , Pages 86–112, https://doi.org/10.1093/bib/bbk007

[19] Ben-Naim, A., 2017, "Information Theory", World Scientific.

[20] Yockey, H.P., 1992, "Information theory and molecular biology", Cambridge University Press.

[21] Yockey, H.P., 2006, "Information Theory and the Origin of Life", Cambridge University Press, .

[22] Witten, I.H., 2017, Frank, E., Hall, M.A., C.J Pal, *Data Mining*, Morgan Kaufmann Cambridge, Ma .

[23] Huang, K., "Statistical Physics and Protein Folding", World Scientific (2005).

[24] Yang, L. "Fighting Chaos with Chaos in Lasers", Science 361, 1201 (2018)

[25] Cao, H., 1999, Zhao, Y.G., Ho, S.T., Seelig, E.W., Wang, Q.H., and Chang, R.P.H., "Random Laser Action in Semiconductor Powder", Physical Review Letters 82, 2278-2281.

[26] Chung, H.S., 2018, & Eaton, W.A. ,"Protein folding transition path times from single molecule FRETG", Current Opinion in Structural Biology, 48, 30-39.

[27] Presse, S., 2013, Ghosh, K., Lee, J. and Dill, K., "Principles of maximum entropy and maximum caliber in statistical physics", Reviews of Modern Physics, 85, 1115-1141.